\documentclass[reprint,amsmath,amssymb,superscriptaddress]{revtex4-2}
\usepackage{braket}% Bras and Kets
\usepackage{graphicx}% Include figure files
\usepackage{dcolumn}% Align table columns on decimal point
\usepackage{bm}% bold math
\usepackage{color} % highlighted text
\usepackage{ulem}
\usepackage{xcolor}

\begin{document}
\title{Second-scale rotational coherence and dipolar interactions \\ in a gas of ultracold polar molecules}  
\author{Philip D. Gregory}
\email{p.d.gregory@durham.ac.uk}
\address{Joint Quantum Centre (JQC) Durham-Newcastle, Department of Physics, Durham University, Durham, United Kingdom, DH1 3LE.}
\author{Luke M. Fernley}
\address{Joint Quantum Centre (JQC) Durham-Newcastle, Department of Physics, Durham University, Durham, United Kingdom, DH1 3LE.}
\author{Albert Li Tao}
\address{Joint Quantum Centre (JQC) Durham-Newcastle, Department of Physics, Durham University, Durham, United Kingdom, DH1 3LE.}
\author{Sarah~L.~Bromley}
\address{Joint Quantum Centre (JQC) Durham-Newcastle, Department of Physics, Durham University, Durham, United Kingdom, DH1 3LE.}
\author{\mbox{Jonathan Stepp}}
\address{Department of Physics and Astronomy, Rice University, Houston, Texas 77005, USA.}
\author{Zewen Zhang}
\address{Department of Physics and Astronomy, Rice University, Houston, Texas 77005, USA.}
\author{Svetlana Kotochigova}
\address{Department of Physics, Temple University, Philadelphia, Pennsylvania 19122, USA.}
\author{Kaden R. A. Hazzard}
\address{Department of Physics and Astronomy, Rice University, Houston, Texas 77005, USA.}
\address{Rice Center for Quantum Materials, Rice University, Houston, Texas 77005, USA.}
\author{Simon L. Cornish}
\email{s.l.cornish@durham.ac.uk}
\address{Joint Quantum Centre (JQC) Durham-Newcastle, Department of Physics, Durham University, Durham, United Kingdom, DH1 3LE.}

\begin{abstract}
Ultracold polar molecules uniquely combine a rich structure of long-lived internal states with access to controllable long-range, anisotropic dipole-dipole interactions. In particular, the rotational states of polar molecules confined in optical tweezers or optical lattices may be used to encode interacting qubits for quantum computation or pseudo-spins for simulating quantum magnetism. As with all quantum platforms, the engineering of robust coherent superpositions of states is vital. However, for optically trapped molecules, the coherence time between rotational states is typically limited by inhomogeneous differential light shifts. Here we demonstrate a rotationally-magic optical trap for RbCs molecules that supports a Ramsey coherence time of 0.78(4)\,seconds in the absence of dipole-dipole interactions. This extends to $>1.4$\,seconds at the 95\% confidence level using a single spin-echo pulse. In our magic trap, dipolar interactions become the dominant mechanism by which Ramsey contrast is lost for superpositions that generate oscillating dipoles. By changing the states forming the superposition, we tune the effective dipole moment and show that the coherence time is inversely proportional to the strength of the dipolar interaction. Our work unlocks the full potential of the rotational degree of freedom in molecules for quantum computation and quantum simulation.
\end{abstract}

\maketitle

The rotational states of polar molecules, together with their controllable dipole-dipole interactions, may be used to encode and entangle qubits~\cite{DeMille2002,Yelin2006,Pellegrini2011,Wei2016,Ni2018,Hughes2020,Zhang2022,Wang2022,Asnaashari2023}, qudits~\cite{Sawant2020}, pseudo-spins~\cite{Barnett2006,Micheli2006,Capogrosso-Sansone2010,Pollet2010,Gorshkov2011,Gorshkov2011b,Zhou2011,Hazzard2013,Zoller2013}, or synthetic dimensions~\cite{Sundar2018,Sundar2019,Feng2022,Cohen2022}. So far, this capability has been exploited to study \mbox{spin-1/2} XY models in a range of geometries~\cite{Yan2013,Seesselberg2018,Bao2022,Holland2022,Li2023,Christakis2023}, and to engineer iSWAP gates that prepare pairs of tweezer-confined molecules in maximally-entangled Bell states~\cite{Bao2022,Holland2022}. Such experiments rely upon the precise control of molecule position that comes from using optical lattices and tweezer arrays. However, spatially-varying and state-dependent light shifts in these traps generally produce a dominant source of decoherence, severely restricting the duration of coherent quantum dynamics. 

`Magic-wavelength' traps have been an invaluable tool in engineering atomic~\cite{Takamoto2005} and molecular~\cite{Kondov2019, Leung2023} clocks that are insensitive to light shifts. The general method being to choose a trap wavelength such that the polarisability of the target states are the same. However, achieving long coherence for rotational states in ultracold molecules has proved difficult, due to the anisotropic interaction with the trap light. The resulting differential light shifts lead to unwanted shifts in the frequency of the rotational transition across the trap. The only implementation of a magic-wavelength trap for rotational transitions has been in fermionic $^{23}$Na$^{40}$K molecules~\cite{Bause2020}. In this case, coherence was limited to $\sim1$\,ms by inhomogeneities in the dc electric field that was also required as part of the scheme. Recent experiments using $^{23}$Na$^{87}$Rb molecules in a near-magic optical lattice reported single-particle rotational coherence times of 56(2)\,ms \cite{Christakis2023}. Other attempts to produce rotationally-magic traps have sought to match polarisabilities by tuning either the polarisation~\cite{Kotochigova2010, Neyenhuis2012, Seesselberg2018, Burchesky2021, Tobias2022,Park2023} or intensity~\cite{Blackmore2018} of the trap light. Here however, residual differential light shifts may still occur due to hyperfine couplings that are quadratic in intensity~\cite{Blackmore2020}. Microwave pulse sequences can be designed to minimise the effects of single-particle dephasing resulting from small residual light shifts or electric field inhomogeneity, for example. Most notably spin-echo~\cite{Yan2013,Seesselberg2018} or XY8~\cite{Bao2022,Li2023} sequences have been used. To date, the longest rotational coherence time reported without rephasing is 93(7)\,ms for single CaF molecules confined to optical tweezers with the polarisation set to a magic angle; this was extended to 470(40)\,ms using a spin-echo sequence~\cite{Burchesky2021}.

%Article, 3000 words. 
In this article, we report second-scale rotational coherence times in a dilute gas of optically-trapped $^{87}$Rb$^{133}$Cs molecules (hereafter RbCs). We engineer a magic-wavelength trap by tuning the frequency of the trap light in the vicinity of a nominally forbidden molecular transition. We observe a Ramsey coherence time of 0.78(4)\,s that is limited primarily by the stability of the trap laser frequency in a state configuration without dipole-dipole interactions. Introducing a single spin-echo pulse, we observe no loss of rotational coherence over 0.7\,s and estimate a minimum coherence time of $>1.4$\,s at the 95\% confidence level. We show that under these conditions, dipolar interactions become the dominant source of decoherence for superpositions that generate oscillating dipoles. We control the strength of these interactions by changing the states forming the superposition. We demonstrate that the coherence time is inversely proportional to the strength of the resonant dipole-dipole interactions.

\begin{figure*}
\includegraphics[width=1.055\textwidth]{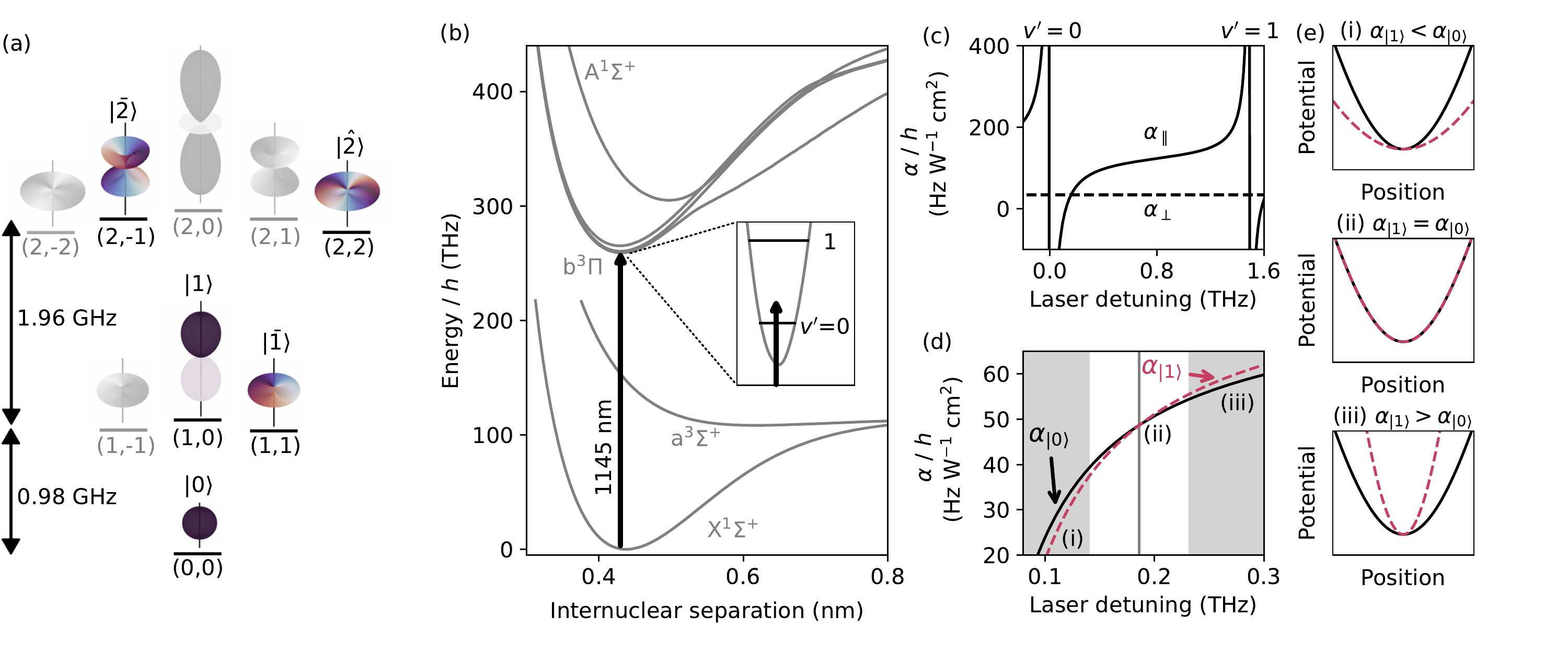}
\vspace{-1cm}
\caption{{\bf A rotationally-magic trap for ultracold molecules.} (a)~Relevant rotational states in this work labelled by $(N, M_N)$. Wavefunctions for each state are shown with phase information for the states used in this work represented by the color. (b)~Electronic structure of RbCs, with the energy corresponding to the 1145\,nm wavelength of the trap laser indicated by the vertical black arrow. (c)~By tuning the laser frequency between the transitions to $v'=0$ and $v'=1$ vibrational levels of the $\mathrm{b}^3\Pi$ potential we vary only the component of the polarisability parallel to the internuclear axis of the molecule $\alpha_\parallel$ whilst keeping the perpendicular component $\alpha_\perp$ constant. (d)~Polarisability for states $\ket{0}$ and $\ket{1}$ as a function of laser detuning from the transition to $\mathrm{b}^3\Pi(v'=0)$. At a detuning of 0.186\,THz the trap is rotationally magic and the polarisability for both states is the same. (e)~Schematic of the relative trap potential for laser detunings such that (i)~$\alpha_{\ket{0}}<\alpha_{\ket{1}}$, (ii)~$\alpha_{\ket{0}}=\alpha_{\ket{1}}$, (iii)~$\alpha_{\ket{0}} > \alpha_{\ket{1}}$.} 
\label{fig:introduction}
\end{figure*}

We start by preparing a thermal gas of ultracold RbCs molecules in their lowest rotational state in an optical trap (see Methods). We typically produce around $\sim2400$ molecules at a temperature of $1.5$\,$\mu$K, and estimated peak density of $6\times10^{10}$\,cm$^{-3}$. We use resonant microwave fields that couple to the molecule-frame dipole moment $d_0=1.23$\,D~\cite{Molony2014} to coherently transfer the molecules between the rotational states shown in Fig.~\ref{fig:introduction}(a). We label the states used in this work by $\ket{0}\equiv(N=0, M_N=0)$, $\ket{1}\equiv(1,0)$, $\ket{\bar{1}}\equiv(1,1)$, $\ket{\bar{2}}\equiv(2,-1)$, and $\ket{\hat{2}}\equiv(2,2)$. Here, $N$ describes the rotational angular momentum, and $M_N$ denotes the dominant projection along the quantisation axis. All these states have the same dominant nuclear spin projections of $m_\mathrm{Rb}=3/2$ and $m_\mathrm{Cs}=7/2$~(see Supplementary Section\,I).

Optical trapping relies upon light with intensity~$I$ interacting with the dynamic polarisability of the molecule~$\alpha$, such that there is a perturbation in energy~$-\alpha I / (2\epsilon_0 c)$ where $\epsilon_0$ is the permittivity of free space and $c$ is the speed of light. Because diatomic molecules are not spherically symmetric, the polarisability along the internuclear axis, $\alpha_\parallel$, is different from that perpendicular to the axis, $\alpha_\perp$, with the two polarisabilities arising from electronic transitions in the molecule with different symmetries~\cite{Kotochigova2006,Vexiau2017}. This results in a molecular polarisability that depends on the orientation of the molecule and can be separated into an isotropic $\alpha^{(0)}$ and anisotropic $\alpha^{(2)}$ component such that \mbox{$\alpha(\theta)=\alpha^{(0)}+\alpha^{(2)}(3\cos^2\theta-1)/2$}. Here, $\theta$ is the angle of the laser polarisation with respect to the internuclear axis, $\alpha^{(0)}=(\alpha_\parallel+2\alpha_\perp)/3$ and $\alpha^{(2)}=2(\alpha_\parallel-\alpha_\perp)/3$. The presence of the anisotropic component leads to a polarisability and therefore light shifts that are dependent on the rotational angular momentum $N$, the projection along the quantisation axis $M_N$, and the angle between the trap laser polarisation and the quantisation axis \cite{Gregory2017}.

To produce a rotationally magic trap we tune the value of $\alpha^{(2)}$ to be zero. This is achieved by trapping with light at a wavelength of \mbox{$1145$\,nm} following a scheme proposed by Guan~$et~al.$~\cite{Guan2021}. We tune the laser frequency to be between transitions to the $v'=0$ and $v'=1$ vibrational states of the mixed $\mathrm{b}^3\Pi$ potential, as indicated in Fig.~\ref{fig:introduction}(b). Transitions to this potential are nominally forbidden from the $\mathrm{X}^1\Sigma^+$ ground state, but may be driven due to weak mixing of $\mathrm{b}^3\Pi$ with the nearby $\mathrm{A}^1\Sigma^+$ potential. Coupling to $\mathrm{A}^1\Sigma^+$ components allows $\alpha_\parallel$ to be tuned by varying the frequency of the trapping light, with poles in the polarisability occurring for each vibrational state in the $\mathrm{b}^3\Pi$ potential, as shown in Fig.~\ref{fig:introduction}(c). Meanwhile, $\alpha_\perp$ remains nearly constant as the light is red-detuned by $\sim100$\,THz from the bottom of the nearest $^1\Pi$ potential. By setting $\alpha_\parallel = \alpha_\perp$ with the laser frequency, the polarisability of the molecule becomes isotropic such that $\alpha^{(2)}=0$ and $\alpha^{(0)}=\alpha_{\perp}$. In Fig.~\ref{fig:introduction}(d,e) we show the effect of tuning the laser frequency on the optical potentials experienced by molecules in states $\ket{0}$ and~$\ket{1}$. The magic condition where the polarisability, and therefore the optical potential, is the same for molecules in either state occurs at a laser detuning of~$\sim186$\,GHz from the transition to the $v'=0$ state. 

\begin{figure}
\includegraphics[width=0.5\textwidth]{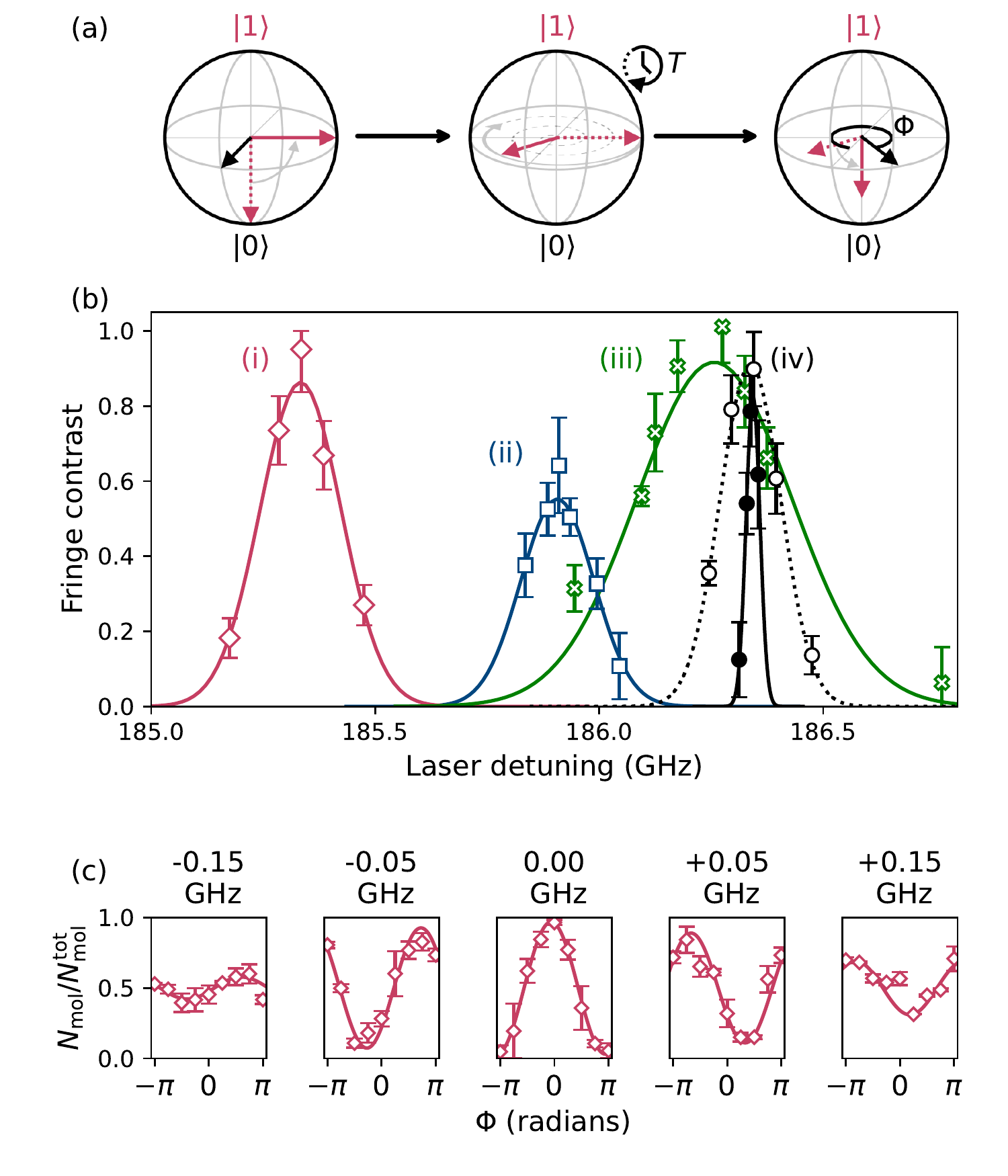}
\vspace{-0.5cm}
\caption{{\bf Optimising coherence time in the magic trap.} (a)~Bloch sphere representation of the Ramsey interferometry sequence. For each step the dotted and solid red arrows represent the initial and final Bloch vector respectively. Solid black arrows indicate the axis about which the Bloch vector is rotated using coherent $\pi/2$ pulses performed on microwave transitions between the neighbouring rotational states.
In the example shown, a $\pi/2$ pulse first prepares the molecules in an equal superposition of states $\frac{1}{\sqrt{2}}(\ket{0}+\ket{1})$. This is then allowed to freely evolve for a time, $T$. 
Finally a second $\pi/2$ pulse with variable phase $\Phi$ is used to project back onto the state $\ket{0}$  for detection. Ramsey fringes are observed as a variation in molecule number $N_\mathrm{mol}$ in state $\ket{0}$ as a function of~$\Phi$. 
(b)~Fringe contrast as a function of trap laser detuning from the transition to $\mathrm{b}^3\Pi(v'=0)$ for state combinations and Ramsey times \mbox{(i)~$\frac{1}{\sqrt{2}}(\ket{0}+\ket{1})$, $T=20$\,ms;} \mbox{(ii)~$\frac{1}{\sqrt{2}}(\ket{0}+\ket{\bar{1}})$, $T=30$\,ms;} \mbox{(iii)~$\frac{1}{\sqrt{2}}(
\ket{1}+\ket{\bar{2}})$, $T=30$\,ms.} Results for the combination \mbox{(iv)~$\frac{1}{\sqrt{2}}(\ket{0}+\ket{\hat{2}})$} are shown for Ramsey times of $T=40$\,ms (empty circles) and $T=175$\,ms (filled circles). The lines show Gaussian fits to each of the results to identify the magic detuning. (c)~Example Ramsey fringes for case (i); the molecule number detected in state $\ket{0}$ is plotted as a fraction of the total number $N_\mathrm{mol}^\mathrm{tot}$.}
\label{fig:optimisation}
\end{figure}

To identify the magic detuning experimentally, we perform Ramsey interferometry as shown schematically in Fig.~\ref{fig:optimisation}(a). For a given pair of states, we fix the Ramsey time and measure the contrast of a Ramsey fringe as a function of the laser detuning (see Methods). We observe a peak in the fringe contrast when the trap light is tuned to be magic as shown in Fig.~\ref{fig:optimisation}(b,c), indicating that the coherence time for that particular combination of states has been maximised. There is a small $\sim1$\,GHz variation in the magic detuning that depends upon the states chosen and the polarisation of the trap light; this is due to the light coupling to different rotational levels of the excited vibrational states~\cite{Guan2021,Fernley2023}. The width of the feature we observe depends on the sensitivity of the differential light shift to the laser frequency, and is inversely proportional to the Ramsey time used.

\begin{figure}
\includegraphics[width=0.5\textwidth]{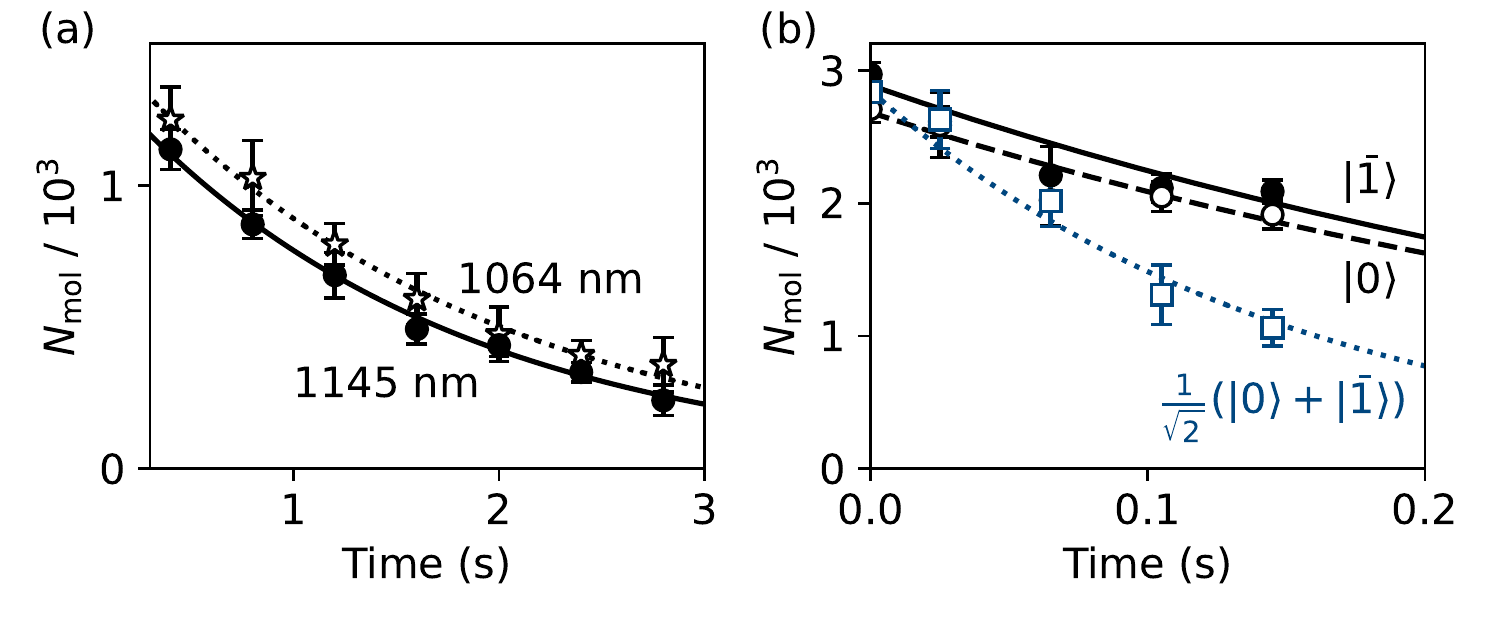}
\caption{{\bf Collisional loss of molecules from the trap.} (a)~Comparison of molecule loss in the magic trap (filled squares) compared to an equivalent trap using light with wavelength of $1064$\,nm (empty stars). (b)~Comparison of loss from the trap for molecules prepared in either $\ket{0}$ or $\ket{\bar{1}}$ with the superposition $\frac{1}{\sqrt{2}}(\ket{0}+\ket{\bar{1}})$. Exponential fits are shown to all results.}
\label{fig:collisions}
\end{figure}

%Collisions 1064/1145
Tuning close to a molecular transition to access a magic wavelength could potentially lead to loss of molecules due to photon scattering. However, we find that our method is compatible with long trap lifetimes. To estimate the scattering rate due to the 1145\,nm light we examine loss of molecules prepared in~$\ket{0}$ from the trap. We begin our measurement after a hold time in the trap of 0.4\,s such that the density of molecules is relatively low and collisional losses~\cite{Gregory2019,Gregory2020} are therefore reduced. We compare the loss from the magic-wavelength trap with loss observed when the wavelength of the trap light is changed to $1064$\,nm, with the intensity set such that the molecules experience the same trap frequencies. The results of both measurements are shown in Fig.~\ref{fig:collisions}(a), with fits from a model assuming exponential decay (see Supplementary Section\,II). We observe similar loss rates, corresponding to lifetimes on the order of $\sim1$\,s in both traps. Assuming the photon scattering rate in the 1064\,nm trap is negligible, we estimate an upper limit on the photon scattering rate in the 1145\,nm trap of $<0.23$\,s$^{-1}$ at the 95\% confidence level. In other work~\cite{Fernley2023}, we have characterised the linewidths of the relevant transitions, with the closest having linewidths $\Gamma_{v'=0}=3.7(4)$\,kHz and $\Gamma_{v'=1}=2.4(3)$\,kHz. Therefore, the trap light is effectively far detuned, with the ratio of the laser detuning to the linewidth of the nearest transition $\Delta/\Gamma_{v'=0}\approx 5 \times 10^7$. It follows that loss due to photon scattering is not an issue for our magic-wavelength trap.

%Dipolar collisions
When molecules are prepared in superpositions of rotational states that are connected by dipole-allowed transitions, they exhibit an oscillating dipole moment in the lab frame. The resultant dipole-dipole interactions can significantly affect the rate of collisional loss of molecules from the trap~\cite{Gregory2019}. In Fig.~\ref{fig:collisions}(b) we compare the loss from the magic trap as a function of time for molecules prepared in either $\ket{0}$, $\ket{1}$ or the superposition $\frac{1}{\sqrt{2}}(\ket{0}+\ket{\bar{1}})$. For the dipolar superposition, we observe a loss rate that is $2.5\times$ greater than for molecules prepared in a single rotational state. The interrogation time available for dipolar samples of molecules is therefore significantly shorter than for non-interacting samples. 

\begin{figure*}
    \centering
    \includegraphics[width=\textwidth]{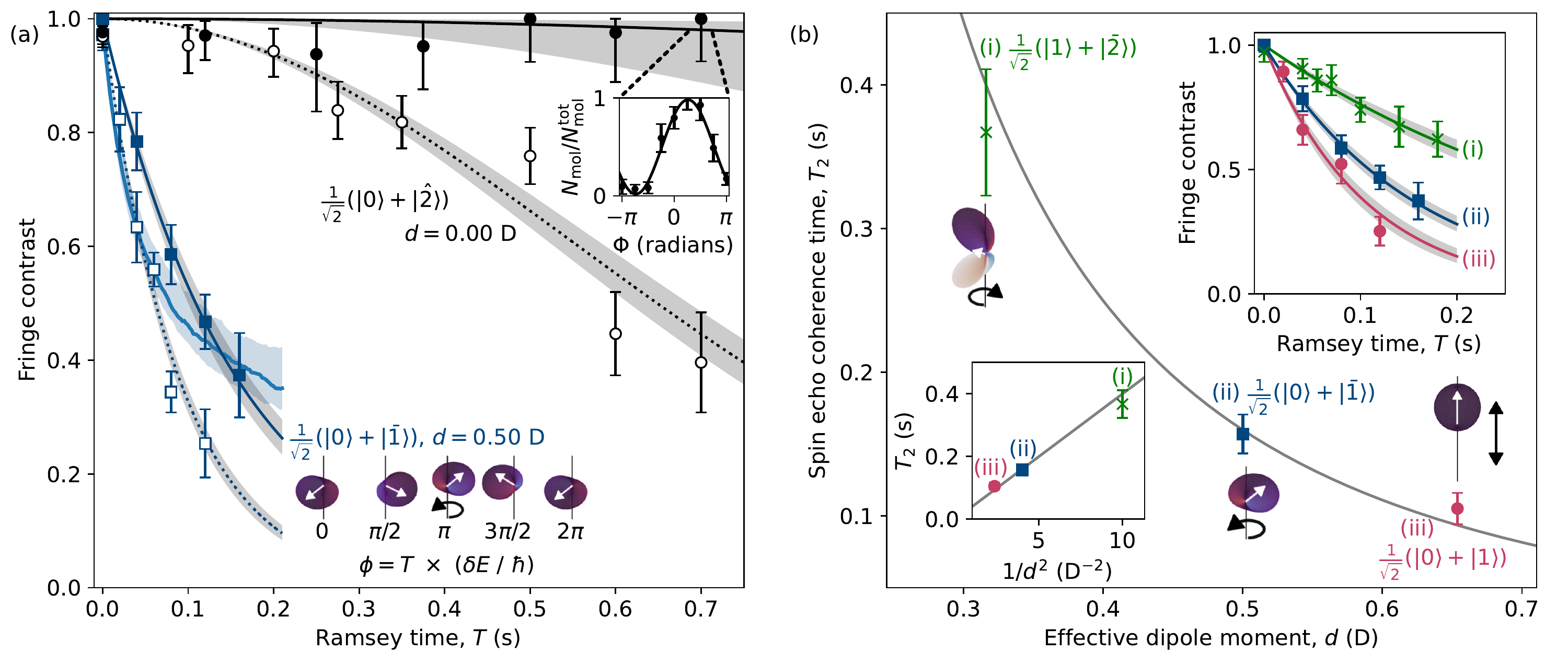}
    \vspace{-0.5cm}
    \caption{{\bf Rotational coherence and dipolar interactions in the magic trap.} (a)~Fringe contrast as a function of the Ramsey time for non-interacting $\frac{1}{\sqrt{2}}(\ket{0}+\ket{\hat{2}})$ (black circles) and dipolar $\frac{1}{\sqrt{2}}(\ket{0}+\ket{\bar{1}})$ (blue squares) superpositions. Empty markers indicate measurements using a standard Ramsey sequence, and filled markers indicate measurements performed with the addition of a single spin-echo pulse. The non-interacting results are fitted using a Gaussian model for decoherence, and the dipolar-results fitted assuming an exponential decay in fringe contrast. The blue lines indicate the decay in fringe contrast from MACE simulations as described in the main text. Uncertainties in the fits and simulations are shown by the shaded regions. The fringes observed for $\frac{1}{\sqrt{2}}(\ket{0}+\ket{\hat{2}})$ with spin echo at $T=0.7$\,s are shown inset. Also shown inset is the wavefunction for the dipolar superposition $\frac{1}{\sqrt{2}}(\ket{0}+\ket{\bar{1}})$ as a function of time~(phase,~$\phi$). The resultant dipole (white arrow) rotates around the quantisation axis (vertical black line) at a frequency proportional to the difference in energy $\delta E$ between the states. (b)~Coherence time in the presence of dipole-dipole interactions. We plot the $1/e$ coherence time measured with spin echo as a function of the effective lab-frame dipole moment, tuned by changing the states used in the spin-echo sequence. The combinations used are (i)~$\frac{1}{\sqrt{2}}(\ket{1}+\ket{\bar{2}})$; (ii)~$\frac{1}{\sqrt{2}}(\ket{0}+\ket{\bar{1}})$; (iii)~$\frac{1}{\sqrt{2}}(\ket{0}+\ket{1})$. The fringe contrast as a function of time for each state combination is shown in the top right inset. The wavefunctions are illustrated with (i) and (ii) yielding dipoles rotating around the quantisation axis, and (iii) resulting in a dipole that oscillates up and down. The bottom left inset shows the coherence time plotted as a function of the dipole-dipole interaction strength $\propto 1/d^2$. }
    \label{fig:coherence}
\end{figure*}

%Second-scale coherence
We first measure the coherence time for a non-interacting sample of molecules by examining the coherence between $\ket{0}$ and $\ket{\hat{2}}$; these are two rotational states not linked by an electric dipole-allowed transition. To perform Ramsey interferometry on this transition, we use a pulse sequence composed of one-photon $\pi/2$ and $\pi$ pulses on the electric dipole-allowed transitions $\ket{0}\leftrightarrow\ket{\bar{1}}$ and $\ket{\bar{1}}\leftrightarrow\ket{\hat{2}}$ (see Supplementary Section\,III). We measure the contrast of the Ramsey fringes as a function of time, shown by the empty circles in Fig.~\ref{fig:coherence}(a). We fit the results with a Gaussian model for decoherence~\cite{Burchesky2021}, where the fringe contrast $C(t) = \exp[-(T/T^*_2)^2]$, to extract the $1/e$ coherence time. From this, we find a coherence time~$T^*_2=0.78(4)$\,s. 

%Expected sources of decoherence
The coherence time we measure is currently limited by residual differential ac Stark shifts in the trap as a result of the light being slightly detuned from the magic wavelength. We estimate the stability of the trap laser frequency to be $\pm0.76$\,MHz ($\pm1\sigma$) over the duration of a typical Ramsey fringe measurement (approximately 30 minutes). This results in a variation of the transition frequency across each fringe measurement of $\pm0.46$\,Hz, with a corresponding theoretical limit on the observed coherence time of $1.1$\,s. There are additional smaller contributions to decoherence arising from the uncertainty in the magic laser frequency extracted from the optimisation curve in Fig.~\ref{fig:optimisation}(b) with $T=
175$\,ms (limit of 4.3\,s), and from a 10\,MHz frequency difference between the two beams used to produce the crossed trap (limit of 8.3\,s). Details of how these limits are calculated are given in Supplementary Section\,IV. In addition, there is a small differential magnetic moment between the states of $0.0124\,\mu_\mathrm{N}$ that adds an additional limit on the coherence time of 10.6\,s associated with noise in the magnetic field ($\sim10$\,mG). Combining all contributions provides an expected limit on the coherence time of 0.74\,s, in excellent agreement with the measured value. Up to an order of magnitude improvement in the coherence time may be achieved by using a better method of laser frequency stabilisation; for example, referencing the light to a high finesse optical cavity would result in a frequency stability of below 100\,kHz~\cite{Gregory2015}. 

%spin echo
We remove most of the effects of these residual differential light shifts by introducing a single spin-echo pulse in the middle of the Ramsey time; this is an effective $\pi$ pulse between $\ket{0}$ and~$\ket{\hat{2}}$ that reverses the direction of precession around the Bloch sphere, thereby cancelling out contributions to single particle dephasing from static inhomogeneities. The result is shown by the filled circles in Fig.~\ref{fig:coherence}(a). We now observe no loss of fringe contrast over the measurement time of 0.7\,s. We do not measure Ramsey fringes for times longer than this due to loss of molecules from the trap diminishing the signal-to-noise ratio. There is a shift in the phase of the Ramsey fringe as a function of Ramsey time that is quadratic (see Supplementary Section\,V). This may be explained by a small imperfection in the spin-echo rephasing, but does not lead to any appreciable loss of coherence. Fitting to the results using the Gaussian model for decoherence, we estimate a minimum coherence time consistent with our results to be $T_2>1.4$\,s at the 95\% confidence level. This result represents the elimination of all decoherence at the detectable precision of our current experiment.

For superpositions of rotational states that lead to oscillating molecular dipoles, dipolar interactions also cause dynamics of Ramsey contrast, and therefore introduce an additional source of decoherence. The dipole-dipole interactions in the system are described by the Hamiltonian~\cite{Gorshkov2011, Gorshkov2011b, Wall2015}
\begin{equation}
\begin{split}
\hat{H}_{\rm{DDI}}&=\\
&\frac{1}{2}\sum_{i\neq j}\frac{1-3\cos^2\Theta_{ij}}{r_{ij}^3}\left(\hat{d}^{(i)}_0\hat{d}^{(j)}_0 + \frac{\hat{d}^{(i)}_1\hat{d}^{(j)}_{-1} + \hat{d}^{(i)}_{-1}\hat{d}^{(j)}_1}{2} \right)
\label{eq:Dipole}
\end{split}
\end{equation}
where $\hat{d}_0, \hat{d_1}, \hat{d_{-1}}$ are spherical components of the dipole operator, $\Theta_{ij}$ is the angle between the vector connecting two molecules and the quantisation axis, and $r_{ij}$ is the inter-molecular distance. The local spatial configuration of molecules varies across the sample. Moreover as the molecules are not pinned by an optical lattice, their configuration is time dependent due to motion of the molecules around the trap.

%dipolar interactions
We examine the coherence between the states $\ket{0}$ and~$\ket{\bar{1}}$ which are connected via a dipole-allowed transition. An equal superposition of these states produces a dipole that rotates around the quantisation axis with magnitude given by the transition dipole moment $d_0/\sqrt{3}$. However, due to the factor of $2$ in the denominator of the final term of Eq.~\ref{eq:Dipole}, this contributes an effective dipole $d=d_0/\sqrt{6}=0.5$\,D in the lab frame. At the peak densities in our experiments, this corresponds to a typical interaction strength of $\sim h\times2$\,Hz. The Ramsey fringe contrast measured as a function of time is shown in Fig.~\ref{fig:coherence}(a) by the blue squares, with (filled) and without (empty) a spin-echo pulse. We see a dramatic reduction in the coherence time measured using either pulse sequence when compared to the non-interacting case. Moreover, the results are no longer well described by the Gaussian model for decoherence. Instead, we fit the results assuming an exponential decay of fringe contrast $C(t)=\exp(-T/T^{\rm{DDI}}_2)$. We find a $1/e$ coherence time of $89(5)$\,ms without the spin-echo pulse and $T^{\rm{DDI}}_2$=157(14)\,ms using the spin-echo sequence. Note that the residual ac Stark shifts that affect the results without spin echo vary depending on the combination of states; we expect that the uncertainty in the magic detuning is the dominant source of dephasing for this combination as collisional losses and dipolar decoherence force us to use a shorter Ramsey time in the optimisation of the trap laser frequency. However, the difference in coherence time between dipolar and non-interacting samples that is observed with the spin-echo pulse can be attributed to the effect of dipole-dipole interactions alone. 

%varying dipolar interactions
We tune the strength of the dipole-dipole interactions in the sample by using different combinations of states. In Fig.~\ref{fig:coherence}(b) we show the $1/e$ coherence time measured with spin echo for three different combinations of states as a function of their effective lab-frame dipole moments. Here, the dipole moment is varied from 0.31\,D to 0.65\,D, calculated using the full state compositions given in Supplementary Section~I. The laser frequency is set to maximise the coherence time for each state combination. As expected, we see in the inset to Fig.~\ref{fig:coherence}(b) that the coherence time is inversely proportional to the magnitude of the interaction strength $U_{ij}\propto d^2$, which confirms that dipolar interactions are the dominant source of decoherence. Moreover, this result demonstrates application of our magic-wavelength trap to molecules in a range of rotational and hyperfine states, allowing control and tunability of the strength of the dipolar interactions.

We compare the decay of fringe contrast  observed in the experiment to that calculated using the Moving-Average Cluster Expansion (MACE)
method~\cite{Hazzard2014} for molecules fixed in space (see Methods). Losses are included in the theory by assuming molecules are lost at a constant rate independent of other molecules. We use the $1/e$ loss time of 0.14(5)\,s determined from an exponential fit to the experimental results in Fig.\,\ref{fig:collisions}(b); we measure similar loss rates for all three of the dipolar combinations investigated. 
Decreases in density from loss noticeably slow down the dynamics, as shown in Supplementary Fig.~1(a). Without fitting, using only the measured loss rates, densities, and trap parameters, the MACE calculations reproduce the timescale for the decay in the Ramsey fringe contrast, as well as the dependence on the choice of state-pair, and the overall monotonic decrease; the result for $\frac{1}{\sqrt{2}}(\ket{0} + \ket{\bar{1}})$ is shown in Fig.~\ref{fig:coherence}(a).  Some details differ, most noticeably that the MACE results predict modestly but systematically shorter timescales and a more concave dependence on time than the experimental results. The theory contrast is between the measured echo and no-echo contrast, and differs from the echo results by  roughly the same factor for all state-pair choices, suggesting a common underlying cause. Molecule motion during the dynamics is a likely source. The reasonably good agreement of MACE calculations and experiment for the overall timescale of contrast decay, and the dependence of this timescale on the strength of the dipole-dipole interactions, supports the conclusion that dipole-dipole interactions are the main cause of the contrast decay for state combinations that generate oscillating molecular dipoles.

%Discussion + conclusions
In conclusion, we have demonstrated a rotationally-magic trap for RbCs molecules, where the effects of all experimentally relevant sources of decoherence may be suppressed, resulting in a coherence time in excess of $1.4$\,s for non-interacting rotational superpositions. Crucially, the magic wavelength is sufficiently far detuned from neighbouring transitions that we observe negligible photon scattering rates and hence long trap lifetimes. We have shown that this provides unparalleled access to controllable dipole-dipole interactions between molecules. Our approach of trapping using light detuned from the nominally forbidden $\mathrm{X}^1\Sigma(v=0)\rightarrow\mathrm{b}^3\Pi(v'=0)$ transition is applicable to other bialkali molecules~\cite{Bause2020,Guan2021}. 

Our work enables the construction of low-decoherence networks of rotational states, which are the foundation for many future applications of ultracold molecules from quantum computation~\cite{DeMille2002,Yelin2006,Pellegrini2011,Wei2016,Ni2018,Hughes2020,Zhang2022,Sawant2020} and simulation~\cite{Barnett2006,Micheli2006,Capogrosso-Sansone2010,Pollet2010,Gorshkov2011,Gorshkov2011b,Zhou2011,Hazzard2013,Zoller2013,Sundar2018,Sundar2019,Feng2022,Cohen2022}, to precision measurement of fundamental constants~\cite{Klos2022}. 
The next step for experiments will be to construct molecular arrays using light at this magic wavelength. For molecules in optical tweezer arrays, this will enable high-fidelity quantum gates using resonant dipolar exchange, either directly between molecules~\cite{Bao2022,Holland2022} or mediated via Rydberg atoms~\cite{Zhang2022,Wang2022,Guttridge2023}. For molecules in optical lattices, long rotational coherence times can be combined with long lifetimes. For a lattice depth of $20$ recoil energies, we predict a one-photon scattering rate of 0.006\,s$^{-1}$, corresponding to a lifetime in excess of 100\,s. For molecules in the magic-wavelength lattice, nearest neighbours will be separated by $r=573$\,nm, and will experience an interaction strength of $h\times343$\,Hz for the largest effective dipoles explored here (and $\Theta=\pi/2$). This corresponds to a timescale for dipolar spin exchange dynamics of~$2.9$\,ms, far shorter than both the coherence time and the lifetime. Techniques for the production of ordered lattice arrays of ground state RbCs molecules have already been demonstrated~\cite{Reichsoellner2017,Das2023}, and are compatible with a magic-wavelength lattice. Our work therefore unlocks the potential of ultracold molecules in optical lattices for simulating quantum magnetism.

\section{Acknowledgements}
We thank J. M. Hutson for many insightful discussions, and thank S. Ospelkaus and M. Tarbutt for the loan of tapered amplifiers used in the early stages of this project. The Durham authors' work was supported by UK Engineering and Physical Sciences Research Council (EPSRC) Grants EP/P01058X/1, EP/P008275/1 and EP/W00299X/1, UK Research and Innovation (UKRI) Frontier Research Grant EP/X023354/1, the Royal Society and Durham University.  K. R. A. H. acknowledges support from the Robert A. Welch Foundation (C-1872), the National Science Foundation (PHY-1848304), the Office of Naval Research (N00014-20-1-2695), and the W. F. Keck Foundation (Grant No. 995764). The work of S.K. was supported by the US Air Force Office of Scientific Research Grants Nos. FA9550-21-1-0153 and FA9550-19-1-0272, the NSF Grant No. PHY-1908634. 

\renewcommand{\theequation}{M\arabic{equation}}
\section{Methods}
\subsection{Production of ground state molecules}
We produce ultracold RbCs molecules from a pre-cooled mixture of Rb and Cs atoms. The atomic mixture is confined to a crossed optical dipole trap using light with a wavelength of 1550\,nm, with a magnetic field gradient applied to cancel the force due to gravity~\cite{McCarron2011}. To form molecules, we sweep the magnetic field down across an interspecies Feshbach resonance at 197\,G~\cite{Koppinger2014}. We then remove the remaining atoms from the trap by increasing the magnetic field gradient to over-levitate the atoms, following which the magnetic field gradient is removed. With the exception of the measurements shown in Fig.~\ref{fig:collisions}(a), at this point the molecules are transferred to the magic trap by ramping the power in the 1145\,nm light on over 30\,ms, and then the power in the 1550\,nm trap off over a further 5\,ms. Finally, we transfer the molecules to the $\mathrm{X}^1\Sigma$ ground state $\ket{0}$ using stimulated Raman adiabatic passage~\cite{Molony2014,Molony2016}. This final step is performed with the trap light briefly turned off to avoid spatially varying ac Stark shifts of the transitions. For the measurements in Fig.~\ref{fig:collisions}(a), we increase the power in the 1550\,nm trap after the removal of atoms and transfer to the magic trap following the ground state transfer. Throughout all of the measurements shown, the molecules are subject to a fixed 181.5\,G magnetic field, and there is no electric field. To detect molecules in $\ket{0}$, we reverse the association process, breaking the molecules back apart into their constituent atoms, which we detect using absorption imaging.

\subsection{Details of the magic trap}

The magic trap is formed from two beams, each with a waist of 50\,$\mu$m, crossing at an angle of $20^{\circ}$. The beams propagate and are polarised in the plane orthogonal to the applied magnetic field that defines the quantisation axis. Both beams are derived from the same laser (Vexlum VECSEL), so to avoid interference effects, we set a 10\,MHz difference in frequency between the beams. The laser detuning reported in Fig.~\ref{fig:optimisation}(b,c) is the average detuning of the two beams. The intensities of the beams are not actively stabilised, but are monitored to ensure they are passively stable to $<5\%$ variation over the course of each measurement. The typical trap frequencies experienced by ground state molecules at the magic detuning are $[\omega_x,\omega_y,\omega_z]=[29(1),144(5),147(5)]$\,Hz. After 15\,ms in the magic trap, we measure the temperature of the ground-state molecules to be~$1.5(2)\,\mu$K using time-of-flight expansion of the cloud over 6\,ms.

The frequency of the 1145\,nm laser is stabilised using a scanning transfer cavity lock~\cite{Subhankar2019}, that is referenced to a 977\,nm laser that is in turn locked to a high finesse cavity with an ultra-low expansion glass spacer~\cite{Gregory2015}. The lock makes corrections to the laser frequency at a rate of $\sim100$\,Hz. This slow feedback rate, together with the relatively low finesse of the transfer cavity $\sim400$, limits the frequency stability of the laser in the current experiments.

\subsection{Coherent state control}
We use coherent one-photon microwave pulses to perform the Ramsey interferometry, during which the trap light is turned off. The microwave sources are referenced to a 10\,MHz GPS clock. We set the microwaves on resonance with the desired transition and calibrate the duration of the pulses using one-photon spectroscopy as described in~\cite{Gregory2016}. The pulse sequences used in this work are shown schematically in Supplementary Section II.

\subsection{Analysis of Ramsey fringes}

We observe Ramsey fringes as a variation in the molecule number $N_\mathrm{mol}$ detected in state $\ket{0}$ as a function of the phase difference $\Phi$ between the initialisation and read-out microwave pulses. We fit each measurement with the function
\begin{equation}
N_\mathrm{mol}(\Phi) = N_\mathrm{mol}^\mathrm{tot}(1+C\cos(\Phi-\Phi_0))
\end{equation}
where $N_\mathrm{mol}^\mathrm{tot}$ is the total number of molecules in the sample, $\Phi_0$ is the phase offset in the Ramsey fringe, and $C$ is the contrast. 

We use a bootstrap fitting algorithm to estimate the uncertainty in the fringe contrast. For a given fringe measurement, we randomly sample the measured $N_\mathrm{mol}$ for each value of $\Phi$ to build up a new dataset that is the same size as the original. We fit to this randomly resampled data to extract a coherence time. This process is repeated $1000$ times to build up a distribution of fitted coherence times, from which we calculate a standard deviation that represents the uncertainty in the true value.

\subsection{MACE}
We calculate the dynamics of our system using the Moving-Average Cluster Expansion (MACE) method~\cite{Hazzard2014}. Particle locations are randomly sampled from the thermal distribution based on the measured trap parameters, temperature, and particle number, and are assumed to be fixed for all times at their initial positions.  We calculate the dynamics starting from all molecules in the $\ket{\rightarrow}$ state, which is the state (ideally) prepared by the initial $\pi/2$ pulse in the Ramsey spectroscopy sequence.  We simulate the time evolution of the Hamiltonian in Eq.~\ref{eq:Dipole} projected onto the relevant state pair, which is a spin-1/2 dipolar XX model~\cite{Gorshkov2011,Gorshkov2011b}
\begin{equation}
H = \frac{J_\perp}{2}\sum_{i\neq j}^{} \frac{1}{2}\frac{1-3\cos^2 \theta_{ij}}{r_{ij}^3} \left(S^+_i S^-_j+ \text{h.c.}\right), 
\end{equation}
where $\vec{r}_{ij}=\vec{r}_i-\vec{r}_j$ is the distance between molecules $i$ and $j$, $\theta_{ij}$ is the angle between the quantization axis and $\vec{r}_{ij}$, $S^\pm_i$ are raising/lowering operators, and $J_\perp= -\braket{\uparrow | d_1 | \downarrow}^2$ for state pairs with angular momentum projections differing by $\pm 1$, and $J_\perp=2 \braket{\uparrow | d_0 | \downarrow}^2$ for state pairs with the same angular momentum projection. For the \( \frac{1}{\sqrt{2}}(\ket{0} + \ket{\bar{1}}) \) state pair, $|\frac{J_\perp \rho}{2 h}| = 2.26\text{Hz}$ where \( \rho \) is the estimated peak density ($6\times10^{10}$\,cm$^{-3}$). Each simulation is performed to a time just before the second $\pi/2$ pulse and then the expectation of  $S^x = \sum_i S^x_i$ is calculated, which is the same as  the Ramsey contrast after the pulse. The MACE method constructs a cluster for each $S^x_i$ from molecule $i$ and the $N_c-1$ other molecules with the strongest interactions with atom $i$, where $N_c$ is a convergence parameter of the method. We exactly calculate  $\braket{S^x_i(t)}$  of each resulting cluster. To assess convergence, we have compared the dynamics for $N_c = 2,4,6,8,$ and $10$ as shown in Supplementary Fig.~1(b). The results converge quickly with $N_c$ for the simulation times of interest, and $N_c=6$ is used for the results in Fig.~\ref{fig:coherence}. The contrast is expected to be converged within widths of the plotted lines over most of the time regime shown. 
  
The dynamics of the Ramsey contrast is already roughly captured if one ignores particle loss, but the loss has non-negligible quantitative effects, which we include in our calculations shown in the main text.  Molecules leaving the trap  decrease the density, which causes the contrast to decay more slowly. To include this loss in the MACE calculations,  we assume that molecules are independently lost from the trap at a constant rate, consistent with the measured time-dependence of the particle number. We take the loss rate to be 0.14(5)\,s, as determined experimentally. MACE clusters are built based on the particle distribution at time $t=0$ and do not change over time.  For each cluster, whenever a molecule is lost we set the interactions between the lost molecule and the remaining molecules to zero.  To propagate the dynamics after this event, we re-diagonalize the Hamiltonian.  This modestly increases the computational difficulty, but only by a factor of $N_c$ in the worst case (when all molecules are lost during the timescale under consideration). To obtain good statistics, we average together 10 loss trajectories of $\sim 2400$ molecules each to obtain a stable result. This reduces the statistical error between runs to a maximum of 2\% over the time scales we are working with.

The error bars on the theoretical calculations presented in the main text show the result of the experimental uncertainty in the number of molecules, loss rate, and temperature. For particle number uncertainty, we computed the Ramsey contrast decay for the $\pm 1\,\sigma$ measured particle numbers, and did the same for loss rate uncertainty and temperature uncertainty. These uncertainties were added together in quadrature to obtain the error bounds in Fig.~\ref{fig:coherence}. Each of these errors are much larger than the statistical or MACE convergence errors.

\clearpage

\setcounter{equation}{0}
\setcounter{figure}{0}
\setcounter{section}{0}
\renewcommand{\figurename}{Supplementary Fig.}
\renewcommand{\theequation}{S\arabic{equation}}

\setcounter{equation}{0}
\setcounter{figure}{0}
\setcounter{section}{0}
\renewcommand{\figurename}{Supplementary Fig.}
\renewcommand{\theequation}{S\arabic{equation}}

\section*{Supplementary information: Second-scale rotational coherence and dipolar interactions in a gas of ultracold polar molecules}

\section{State composition}

Composition of the states used in this work, given in the uncoupled basis $(N, M_N, m_\mathrm{Rb}, m_\mathrm{Cs})$, are

\vspace{0.2cm}
\begin{tabular}{rl}
\vspace{0.2cm}
$\ket{0} \equiv$ & {\bf 1.000}$\ket{0,0,\frac{3}{2},\frac{7}{2}}$\\ 
\vspace{0.2cm}
$\ket{1} \equiv$ & {\bf 0.924}$\ket{1,0,\frac{3}{2},\frac{7}{2}}$ - 0.370$\ket{1,1,\frac{1}{2},\frac{7}{2}}$\\
\vspace{0.2cm}
& + 0.091$\ket{1,1,\frac{3}{2},\frac{5}{2}}$ \\
\vspace{0.2cm}
$\ket{\bar{1}} \equiv$ & {\bf 1.000}$\ket{1,1,\frac{3}{2},\frac{7}{2}}$ \\
\vspace{0.2cm}
$\ket{\bar{2}} \equiv$ & {\bf 0.934}$\ket{2,-1,\frac{3}{2},\frac{7}{2}}$
 - 0.220$\ket{2,1,-\frac{1}{2},\frac{7}{2}}$\\
\vspace{0.2cm}
& - 0.207$\ket{2,0,\frac{3}{2},\frac{5}{2}}$
 + 0.168$\ket{2,0,\frac{1}{2},\frac{7}{2}}$ \\
\vspace{0.2cm}
& - 0.056$\ket{2,2,-\frac{3}{2},\frac{7}{2}}$
 + 0.055$\ket{2,1,\frac{1}{2},\frac{5}{2}}$\\
\vspace{0.2cm}
& + 0.039$\ket{2,2,-\frac{1}{2},\frac{5}{2}}$
 - 0.005$\ket{2,2,\frac{1}{2},\frac{3}{2}}$\\
\vspace{0.2cm}
& - 0.001$\ket{2,1,\frac{3}{2},\frac{3}{2}}$ 
 - 0.001$\ket{2,2,\frac{3}{2},\frac{1}{2}}$ \\
\vspace{0.2cm}
$\ket{\hat{2}} \equiv$ & {\bf 1.000}$\ket{2,2,\frac{3}{2},\frac{7}{2}}$ \\
\end{tabular}

\noindent Here, each of the coefficients are given to 3 decimal places, and the dominant contribution in each case has its coefficient highlighted in bold. The energies and compositions of the rotational and hyperfine states are calculated with Diatomic-Py~\cite{Blackmore2023}. 

\section{Lifetime measurements in Fig.\,3(a)}
We fit the results in Fig.\,3(a) with the exponential function $N_\mathrm{mol}(t)=N_\mathrm{mol}^\mathrm{init}e^{-kt}$. Here, $N_\mathrm{mol}^\mathrm{init}$ is the number of molecules present at time $t=0$ and $k$ characterises the rate of loss of molecules from the trap. We extract a loss rate of $k_{1145}=0.61(5)$\,s$^{-1}$ with 1145\,nm light, and $k_{1064}=0.56(7)$\,s$^{-1}$ with 1064\,nm light. This is consistent a rate of loss that is not dependent on the wavelength. To estimate the upper limit to the photon scattering rate that is given in the main text, we calculate the difference in these scattering rates $k_{1145}-k_{1064}=0.05(9)$\,s$^{-1}$, assuming no correlation in the uncertainty of the two measurements. At the 95\% confidence level, this indicates that the difference in loss rate must be below 0.23\,s$^{-1}$. This is broadly consistent with the expected single photon scattering rate which we calculate to be 0.4(1)\,s$^{-1}$ from the known linewidths of the transitions~\cite{Fernley2023} and peak laser intensity in our optical trap (20\,kW\,cm$^{-2}$). 

\section{Pulse sequences}
We use various pulse sequences to perform Ramsey interferometry depending on the combination of states targeted. These are shown schematically below. 
\includegraphics[width=0.45\textwidth]{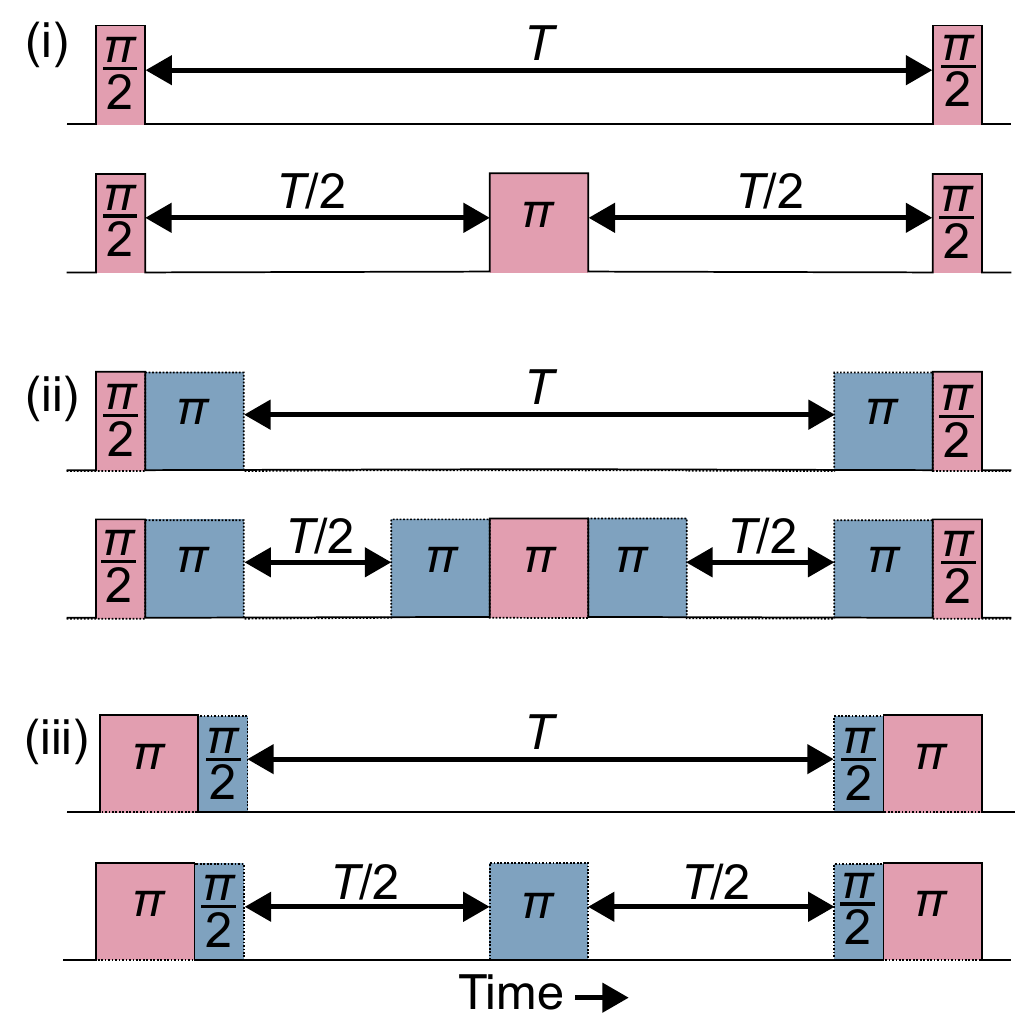}

\noindent At the start of the sequence the molecules are always prepared in $N=0$. The solid line with red fill indicates microwave transitions driven between $N=0$ and $1$ rotational states, and the dotted line with blue fill indicates transitions driven between $N=1$ and $2$. Sequences are used for the combinations (i)~$\frac{1}{\sqrt{2}}(\ket{0}+\ket{1}), \frac{1}{\sqrt{2}}(\ket{0}+\ket{\bar{1}})$ (ii)~$\frac{1}{\sqrt{2}}(\ket{0}+\ket{\hat{2}})$, (iii)~$\frac{1}{\sqrt{2}}(\ket{1}+\ket{\bar{2}})$. In each case the top (bottom) sequence shows the sequence without (with) a spin-echo pulse. To measure a Ramsey fringe, the phase of the last $\pi/2$ pulse is varied. 

\section{Residual light shift contributions to decoherence}
In the absence of dipole-dipole interactions, the coherence time $T_2^*$ is limited by variation $\Delta E$ in the energy difference between two states
\begin{equation}
T_2^*=\frac{h}{|\Delta E|}
\end{equation}
where $h$ is the Planck constant. In Fig.\,4(a), we observe decoherence without interactions which is suppressed by a spin-echo pulse. Here we discuss possible sources of this decoherence, specifically for the superposition $\frac{1}{\sqrt{2}}(\ket{0}+\ket{\hat{2}})$. For our calculations, we assume $\Delta E$ is the $2\sigma$ variation in the transition energy. 

\subsection{Instability in the trap laser frequency}
The trap laser frequency is stabilised by a scanning transfer cavity lock, and we estimate the standard deviation in the laser frequency to be $0.76$\,MHz. For a $2\sigma$ change in the laser frequency, the differential polarisability between $\ket{0}$ and $\ket{2}$ changes by \mbox{$h\times2.9\times10^{-5}$\,Hz\,W$^{-1}$\,cm$^2$}. Moreover, we have directly measured the coefficient relating the light shift to the change in laser frequency to be $6.0(2)\times10^{-7}$ in our trap. The $2\sigma$ variation in the laser frequency therefore corresponds to $2\times(6\times10^{-7})\times0.76\,\mathrm{MHz}=0.92$\,Hz which we assume equal to $\Delta E/h$, yielding a limit on $T_2^*$ of~$1.1$\,s.

\subsection{Uncertainty in optimisation of magic detuning}
We measure the magic detuning for a given state combination as shown in Fig.\,2(b). For the states $\ket{0}$ and $\ket{\hat{2}}$, our most precise measurement of the magic detuning is found using a Ramsey time of $T=175$\,ms, which has $1\sigma$ uncertainty of 3\,MHz. A systematic detuning of 3\,MHz leads to an average light shift experienced by the molecules of $(6\times10^{-7})\times3\,\mathrm{MHz}=1.8$\,Hz. 

Spatial variation in this light shift causes decoherence. We estimate this variation from the known geometry of the trap beam and assuming a thermal cloud of molecules at equilibrium. From this, the $2\sigma$ variation in the light shift experienced will be $13\%$ of the average, i.e. there is spatial variation in the light shift of $0.13\times1.8=0.234$\,Hz. This puts a limit on $T_2^*$ of~4.3\,s.    

\subsection{10\,MHz frequency difference between beams}

There is a 10\,MHz frequency difference between the two beams that form the magic trap in order to eliminate interference effects. When set symmetrically about the magic frequency, there will be a light shift of $(6\times10^{-7})\times \pm 5\,\mathrm{MHz}=\pm3$\,Hz. The effects from each beam are broadly cancelled as the intensities are set to be the same. However, variation in the relative intensities of the beams will vary as molecules move around the trap. We estimate from the known geometry of the trap beams and assuming a thermal cloud of molecules at equilibrium that the $2\sigma$ variation in the beam balance is only~2\%. This leads to variation in the light shift $0.12$\,Hz that corresponds to a limit on $T_2^*$ of 8.3\,s.   

\section{Phase shift as a function of Ramsey time}
We observed a phase shift in the Ramsey fringe as a function of the Ramsey time for non-interacting states with the spin-echo pulse, as discussed in the main text. This phase shift is shown in the graph below, with a quadratic fitted to guide the eye. 
\includegraphics[width=0.45\textwidth]{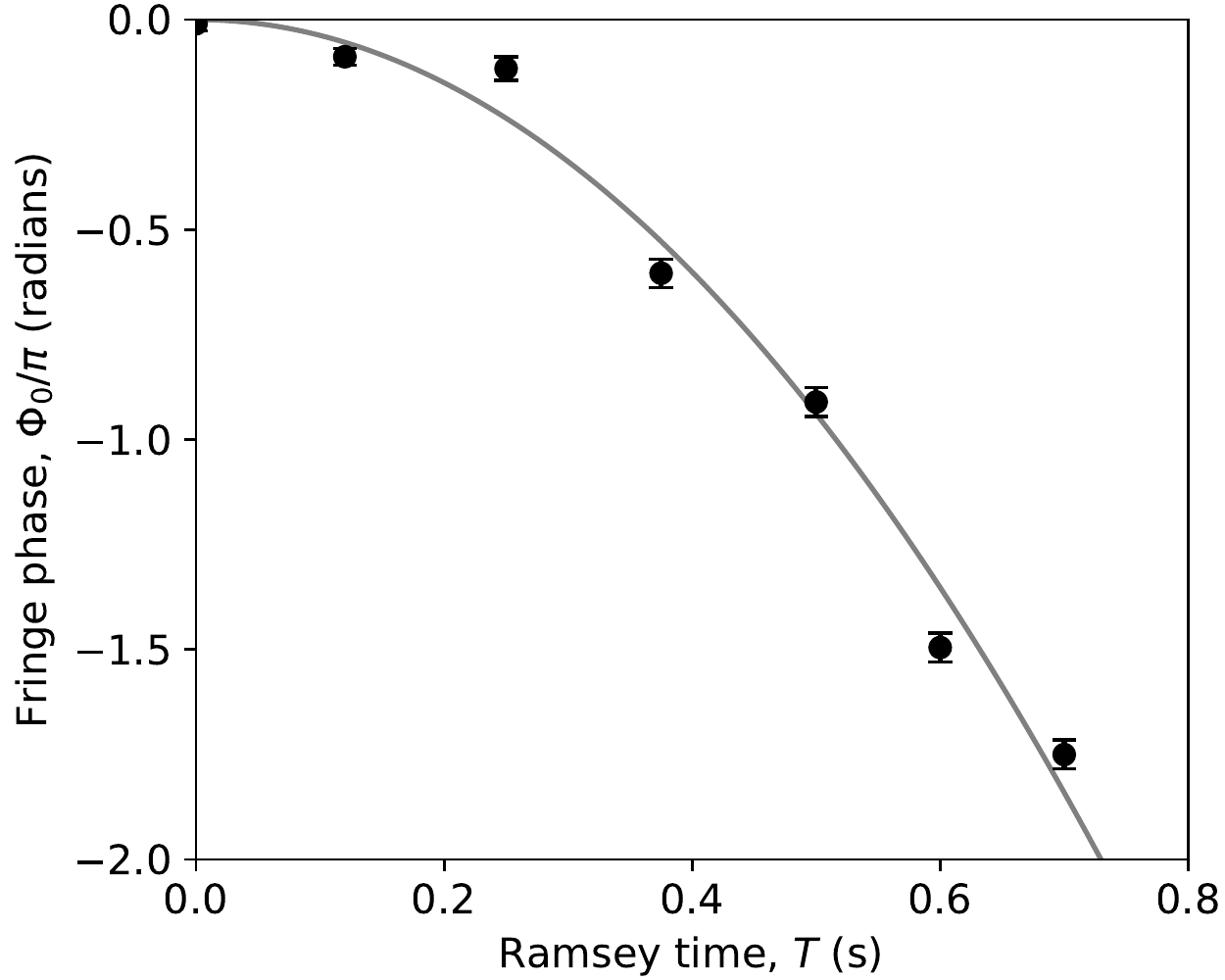}

\begin{figure*}
  \includegraphics[width=\textwidth]{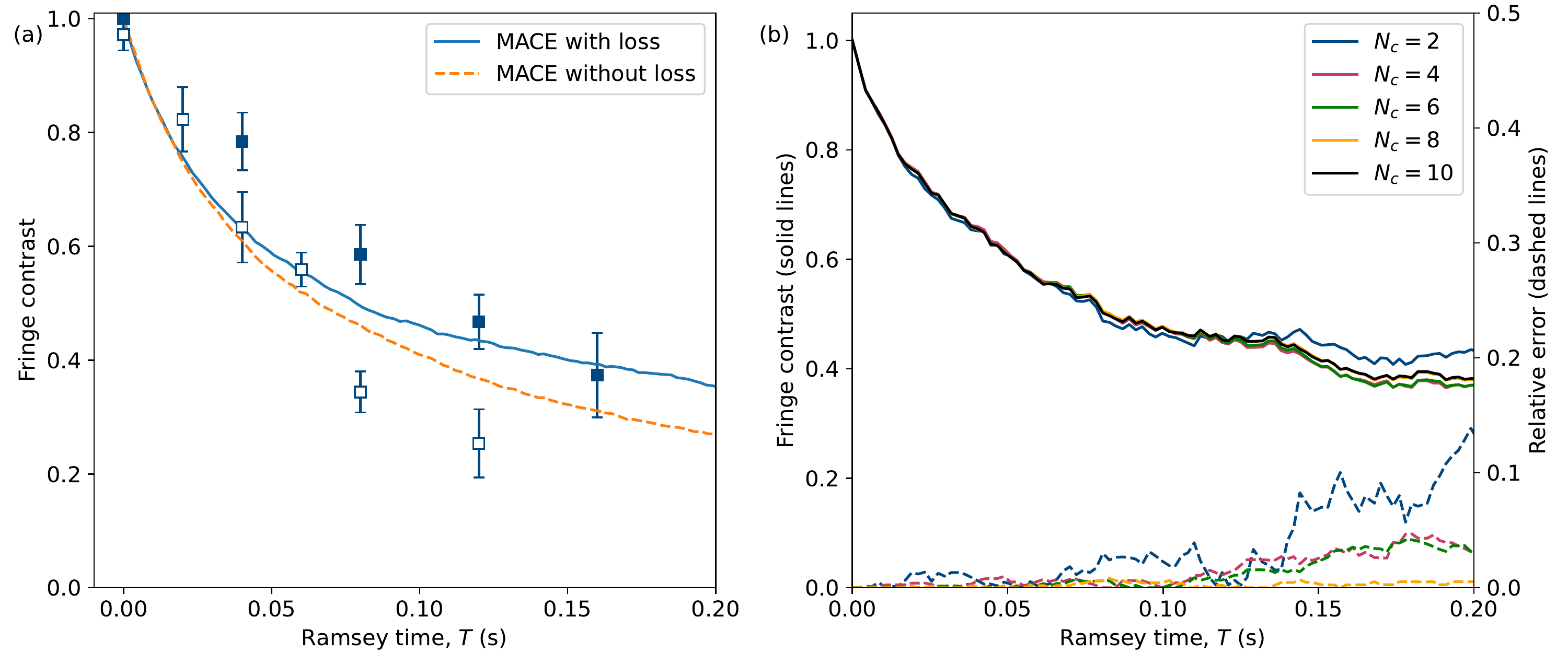}
  \caption{(a)~Comparison of MACE calculation with and without considering loss for the $\frac{1}{\sqrt{2}}(\ket{0} + \ket{\bar{1}})$ state pair. These calculations are both averages over 10 loss trajectories calculated at particle number N = 2391 and particle lifetime 0.137 s. The filled (empty) markers show the experimentally measured fringe contrast with (without) a spin-echo pulse. (b)~Comparison of a single loss trajectory at cluster size \( N_c = 2,4,6,8,10 \) shown by the solid lines, alongside the relative error betwen each MACE run and the \( N_c = 10 \) MACE run shown by the dashed lines.}
  \label{fig:loss_conv}
\end{figure*}

\end{document}